\documentclass [12PT]{article}

\usepackage{graphicx}

\usepackage[dvipsnames,usenames]{color}

\definecolor{DarkGreen}{rgb}{0.5,0.8,0.6}   
\definecolor{RGBblack}{rgb}{0.0,0.0,0.0}    

\definecolor{grau}{rgb}{0.8,0.8,0.8}

\newcommand{\chen}[1]{\color{orange}}

\usepackage{enumitem}
\usepackage{makecell}
\usepackage{booktabs}
\usepackage{subcaption}
\usepackage{blindtext}
\usepackage{algorithm}
\usepackage{longtable}
\usepackage{multicol}
\usepackage{algorithmic}
\usepackage{natbib}
\usepackage{multirow}
\usepackage{bm}
\usepackage{amsthm,amsmath,amssymb}
\usepackage{mathrsfs}
\usepackage{url}
\usepackage[title]{appendix}

\newcommand{\obs}{\text{obs}}
\newcommand{\mis}{\text{mis}}
\newcommand*\samethanks[1][\value{footnote}]{\footnotemark[#1]}

\usepackage{color}

{
	\title{\bf The Backfill i3+3 Design for Dose-Finding Trials in Oncology}

 	\author{
 		Jiaxin Liu\thanks{East China Normal University, Shanghai, CHN} \ \thanks{Cytel Inc., Shanghai, CHN}, Shijie Yuan\samethanks , B. Nebiyou Bekele \thanks{San Ramon, CA, USA}, and Yuan Ji\thanks{Corresponding email:koaeraser@gmail.com; Department of Public Health Sciences, The University of Chicago, Chicago, USA}
 	}
	\date{\today}
}

\begin{document}
\maketitle

\begin{abstract}
  We consider a formal statistical design that allows   simultaneous enrollment of a main cohort and a backfill cohort of patients in a dose-finding trial. The goal is to accumulate more information at various doses to facilitate dose optimization. 
The proposed design, called Bi3+3, combines the simple dose-escalation algorithm in the i3+3 design 
  and a model-based inference under the framework of probability of decisions (POD), both previously published. As a result, Bi3+3 provides a simple algorithm for backfilling patients to lower doses in a dose-finding trial once these doses exhibit safety profile in patients. The POD framework allows dosing decisions to be made when   some backfill patients    are still being followed with incomplete toxicity outcomes, thereby potentially expediting the clinical trial. At the end of the trial, Bi3+3 uses both toxicity and efficacy outcomes to estimate an optimal biological dose (OBD). The proposed inference is based on a  dose-response model that takes into account either a monotone or plateau dose-efficacy relationship, which are frequently encountered in modern oncology drug development. Simulation studies show promising operating characteristics of the Bi3+3 design in comparison to existing designs. 


\end{abstract}

\textbf{\textit{Keywords}}:  Dose Optimization; Efficacy; Phase I; Probability of Decision; Toxicity.

\section{Introduction}
\label{sec:introduction}

In an oncology phase I dose-finding trial, a binary dose-limiting toxicity (DLT) outcome is used as the primary endpoint to assess the toxicity of ascending doses of a new investigation therapeutics. Toxicity is assumed to be monotonically increasing with dose levels. The primary goal is to establish the toxicity profile of the therapeutics and identify the maximum tolerated dose (MTD), which is defined as the highest dose with a toxicity probability no more than a prespecified target rate, say 0.3. Patients are enrolled in cohorts and sequentially treated at different doses. A statistical design is used to guide the sequential patient enrollment and dosing decisions that specify the dose for the next patient cohort. An appropriate design is expected to achieve two goals simultaneously, first to allocate as many patients to safe and potentially efficacious doses and second to quickly learn the safety profile of the therapeutics and identify the MTD. Over the past three decades, a large number of effective statistical designs have been developed that drastically improved the quality and efficiency of phase I trials. 
For example, the continual reassessment method (CRM) \citep{o1990continual} applies model-based inference to estimate the MTD whenever new toxicity outcomes become available. Subsequent model-based designs such as mTPI \citep{ji2010modified} and mTPI-2 \citep{guo2017bayesian} attempt to simplify the statistical models and utilize  up-and-down decision rules to generate simple and transparent decision tables for practical trials. More recent design development aims to further simplify the decision and modeling framework, leading to model-assisted designs like CCD \citep{Ivaccd2007} and BOIN \citep{yuan2015boin}, as well as model-free designs like i3+3 \citep{liu2020i3+}. 

\begin{figure}[h]
  \centering
   \includegraphics[width=0.8\textwidth]{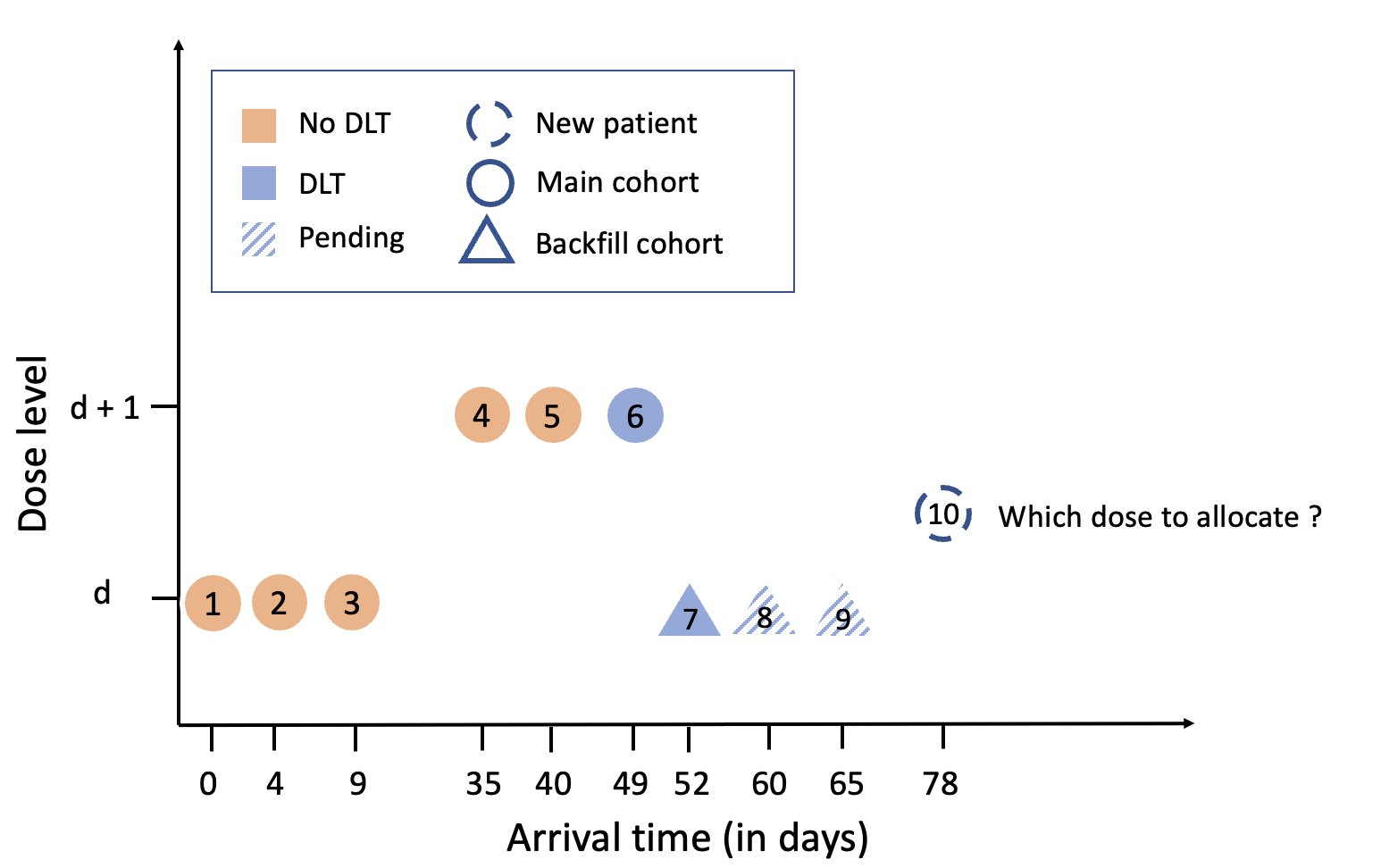}
    \caption{A hypothetical dose-finding trial allowing patient backfill. The main cohort patients are 1-6 and the backfill cohort ones are 7-9. By the time patient 10
arrives, patients 1 - 7 have observed outcomes, and patients 8 and 9 are still being followed without
definitive outcomes. This figure is a stylized example and should not be interpreted to mean that only one dose is allowed for backfill. The  dose for patient 10 is determined by the design algorithm in $\S$ \ref{sec:pat_allocation}.} 
    \label{fig:backfill_problem}
\end{figure}

The FDA Project Optimus \citep{Optimus} advocates for a more thorough and comprehensive exploration of doses in early-phase oncology drug development. Across many new concepts in the project, an overarching message is to encourage designs that gather more information about doses, especially those less toxic but equally efficacious, thereby having a better chance identifying an optimal biological dose in early-phase clinical trials, which is defined as the dose level at which the therapeutic intervention demonstrates maximum efficacy in treating the target condition while minimizing undesirable side effects or toxicity. 
Here, we consider a framework that allows patients to be backfilled to lower doses during a dose-finding trial once   these doses exhibit safety profile in patients.    Backfill is useful in settings where efficacy of a drug does not always increase with dose level, as seen in many recent immune and targeted oncology drugs \citep{shah2021drug}. Therefore, backfilling patients at lower doses provides opportunities to further accumulate information (e.g., pharmacology or efficacy data)  at    these doses while the trial continues to explore higher doses. In this way, efficiency in finding an  optimal dose  can be gained since more information will be available at different doses. However, backfilling raises a new logistic and statistical issue, as illustrated in Figure \ref{fig:backfill_problem}. 
To see this, first note that once backfill cohorts are enrolled   at    a lower dose and the trial continues to explore a higher dose, there are two cohorts of patients that are enrolled and treated at different doses, the backfill cohort at the low dose and the main cohort at the high dose. The main cohort is usually enrolled first after which the backfill cohort is enrolled. When   patient outcomes from the main cohort are recorded,    a dosing decision must be made to   determine    the dose for future patients in the next main cohort. Such decision should utilize the information from the patients in the main cohorts    as well as    the backfill cohorts. However, since patients in backfill cohorts are enrolled after the main cohort, it is possible that, at the time of dosing decision, one or more patients (patients 8 and 9 in Figure \ref{fig:backfill_problem}) in the backfill cohort might still be followed without a definitive DLT outcome. In other words, there could be pending patients with incomplete outcomes in the backfill cohorts. These patients might have been followed for a while but no definitive DLT outcomes can be assessed yet. This raises a question of how the information in these patients could be used in statistical modeling and inference for the dosing decision. One solution is to wait and pause trial enrollment until all the pending patients in the backfill cohort complete follow-up and have their DLT outcomes recorded. The proposed Bi3+3 design aims to provide a better solution to this question that does not always require pausing enrollment.

There are few formal statistical approaches for patient backfill in real-world trials. \citet{manasanch2020phase} used a simple backfill algorithm under the 3+3 design to acquire additional information on low doses during the trial.  \citet{dehbi2021controlled} proposed a novel backfill framework using hypothesis 
testing under a  CRM. The authors propose to test two hypotheses,  one corresponding to the monotonically increasing efficacy model and the other to the plateau efficacy model under the framework of CRM.   They allow a backfill cohort to be  enrolled following every main cohort when the main cohort reaches level 2 or higher. However, Backfill CRM assumes the data in the backfill cohort do not influence the dosing decision for the main cohort.   
Here, we consider a design that is slightly different. We assume that some patients in the backfill cohorts are still being followed without DLT outcomes  and  apply a time-to-event model to incorporate information from these patients  to allow  
dosing decisions to be made with pending outcomes in the backfill cohort.  
Designs that allow dosing decisions in the presence of pending outcomes have been developed for dose-finding trials. For example, TITE-CRM \citep{cheung2000sequential} is a pioneering time-to-event (TITE) model under the CRM framework and has been applied in various trials. Motivated by TITE-CRM, a variety of TITE designs have been developed.  \citet{Barnett2023} incorporate TITE-CRM in a backfill design that is capable of making dosing decisions when there are late onsite toxicities. Recently, a new framework based on probability of decision (POD) has been proposed, represented by the POD-TPI design \citep{zhou2020pod}, to calculate the probability of different decisions by treating the pending outcomes of patients as random  quantities.  The proposed Bi3+3 design adopts the POD framework for the potential pending outcomes due to backfill.

In addition to toxicity data,  efficacy data of phase I trials are collected and should be used for dose selection.  For cytotoxic drugs, the efficacy is assumed to increase with  doses and therefore MTD is  also the optimal biological dose (OBD)   since it gives the highest efficacy among all tolerable doses. However, new targeted and immune therapies are often cytostatic in that the efficacy response rate may reach a plateau and stay flat after certain dose level. When the dose level with highest response is lower than the MTD, the MTD is no longer the OBD, rendering selection of the MTD for future studies suboptimal \citep{shah2021drug}. 
To this end, 
  the proposed  backfill i3+3 design (Bi3+3)  selects an OBD based on both efficacy and toxicity data at the end of the trial. 
  Specifically,  a change-point model is proposed for the dose-efficacy response to capture the potential cytostatic relationship. The remainder of the article is organized as follows. Section \ref{sec:stat_model} introduces the proposed design, and Section \ref{sec:mtd_obd_selection} shows how we select MTD and OBD in the Bi3+3 design. Simulations will be presented in Section \ref{sec:simulation}, and discussion about our work is in the last section, $\S$\ref{sec:discussion}.

\section{Proposed Bi3+3 Design}
\label{sec:stat_model}

\subsection{Notation}
\label{sec:notation}
Assume a total of $D$ ascending doses of a new therapeutics is under investigation. 
At a given moment of the trial, suppose $n_d$ patients have been allocated to dose $d$ among whom $y_d$ experience the DLT, $d=1, \ldots, D.$   Let $p_d$ denote the toxicity probability at dose $d$, and we assume  
$$y_d | n_d,p_d \sim Bin(n_d,p_d).$$
Also assume the efficacy outcome of each patient to be binary which  is  usually observed   10-12 weeks    after the treatment for solid tumors in oncology trials. Letting the number of patients experiencing efficacy be $v_d$ and the response rate $q_d$ at dose $d$, we   assume    
$$v_d | n_d,q_d \sim Bin(n_d,q_d) \qquad d=1,\cdots, D.$$

The proposed Bi3+3 design extends the algorithm in the i3+3 design \citep{liu2020i3+} given in the table below. 
Let $p_T$ denote the target toxicity probability of the MTD, which is the highest toxicity rate that can be tolerated in the trial. Define the equivalence interval (EI) as $[p_T - \epsilon_1, p_T + \epsilon_2]$, which describes a range of toxicity probabilities within which doses are deemed equivalent to the MTD. For example, $p_T = 0.25$ and EI $=[0.2, 0.3].$ Then the dose-finding algorithm of the i3+3 design can be summarised  in Table \ref{tbl:i3+3_algorithm}.  Here, a value  ``below” the EI means that the value is less than $(p_T-\epsilon_1)$, the lower bound of the EI. A value  ``inside” the EI means that the value is larger than or equal to $(p_T-\epsilon_1)$ but less than or equal to $(p_T+\epsilon_2)$, the upper bound of the EI. A value  ``above`` the EI means that the value is larger than $(p_T+\epsilon_2)$. Decisions $E$, $S$, and $D$ represent escalation to the next higher dose, stay at the current dose, and de-escalation to the next lower dose, respectively. 

 \begin{table}[!htbp]
       \begin{center}
        \caption{The  decision rules in the i3+3 design. Notation: $d$ represents the current dose being investigated in the trial; $n_d$ and $y_d$ denote the number of patients enrolled and those with DLT at dose $d$, respectively. }
        \label{tbl:i3+3_algorithm}
	\begin{tabular}{|p{8cm} |{c}| c|}
		\hline
		{\it Condition} & {\it Decision} & {\it Dose for next cohort}\\
		\hline 
		$\frac{y_d}{n_d}$ below EI	& Escalation ($E$)  & $d+1$ \\ \hline
		$\frac{y_d}{n_d}$ inside EI	& Stay ($S$) & $d$ \\ \hline
		$\frac{y_d}{n_d}$ above EI and $\frac{y_d-1}{n_d}$ below EI	& Stay ($S$) & $d$ \\ \hline
		$\frac{y_d}{n_d}$ above EI and $\frac{y_d-1}{n_d}$ inside EI	& De-escalation ($D$) & $d-1$\\ \hline
		$\frac{y_d}{n_d}$ above EI and $\frac{y_d-1}{n_d}$ above EI	& De-escalation ($D$)  & $d-1$\\
		\hline
	\end{tabular}
 \end{center}
\end{table}

Phase I oncology trials enroll patients in cohorts, say a group of three patients   per cohort.    After a cohort is enrolled and assigned to a dose level for treatment, the patients are followed for   three to four weeks    to evaluate drug safety and record any DLT outcomes. While the patients are being followed, new patients may be eligible for trial enrollment. The proposed Bi3+3 design allocates these patients, as backfill cohorts, to lower doses, which have already been tested and exhibit sufficient safety. 

 Assume dose $d$ is currently used for treating patients in the trial. We denote ``main cohort" (mc) the cohort of patients for dose escalation, and ``backfill cohort" (bc) the cohort of patients for backfill at lower doses. At a given moment of the trial, assume   an    mc of patients is allocated to dose $d$. 
 After the mc is allocated, Bi3+3 allocates bc to one or multiple doses  that are  lower than the current dose $d.$ Denote the set of doses assigned to bc by  ${\cal B}_d = \{k_0, k_0 + 1, \cdots, d-1\}$ where 
 $k_0$ in ${\cal B}_d$ is the lowest dose available for backfill.  The proposed Bi3+3 design continuously and adaptively determine $k_0$   throughout the trial.   
 We will describe a method to determine $k_0$ later in $\S$\ref{sec:update_d0}.     But first,  we describe the main algorithm for the Bi3+3 design next. 

\subsection{Patient allocation for Bi3+3}
\label{sec:pat_allocation}
\begin{figure}[!htbp]
    \centering
    \includegraphics[width=0.9\textwidth]{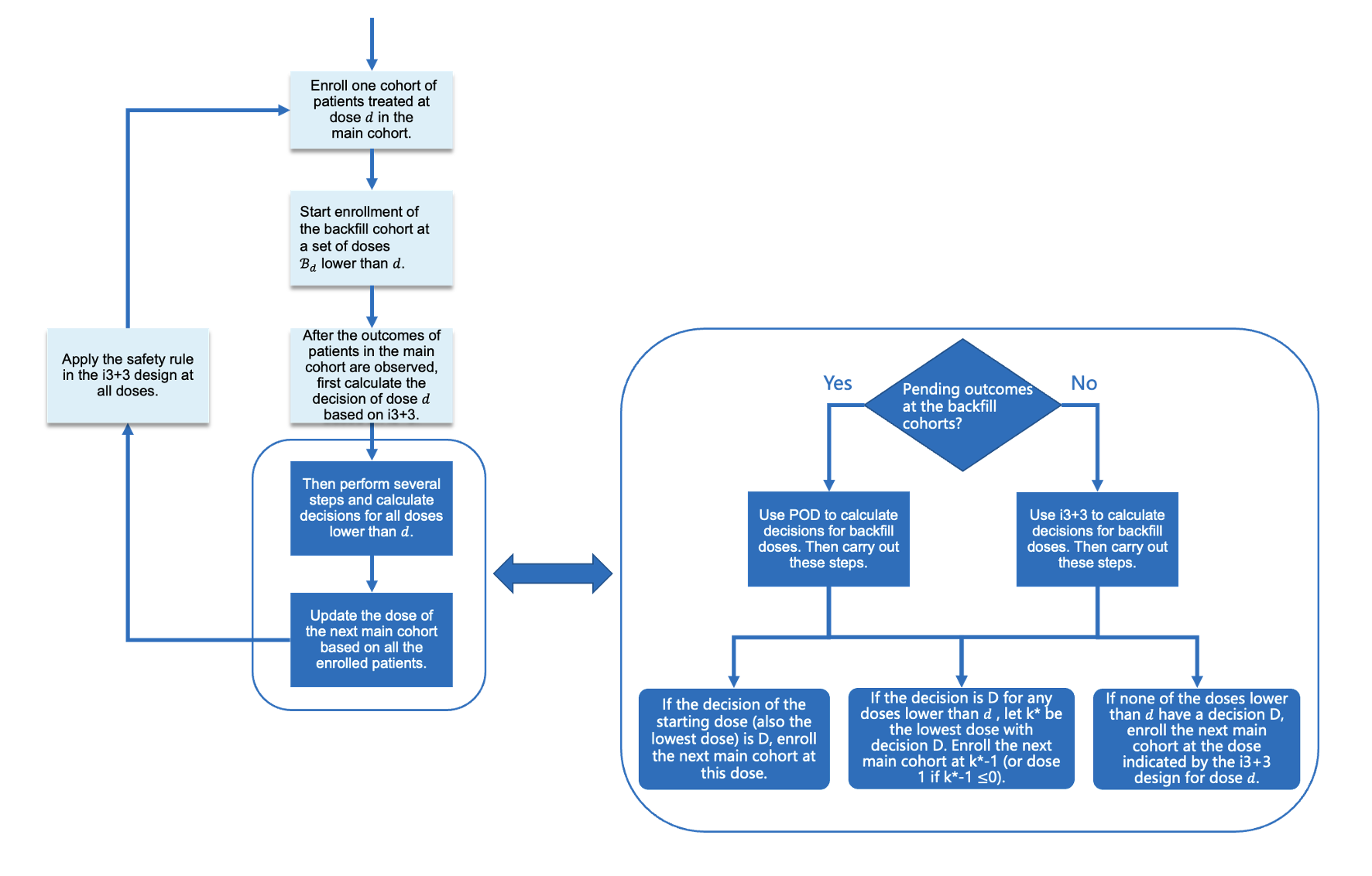}
    \caption{Design flow of Bi3+3 for decision making and patient allocation. } 
    \label{fig:bi3+3_process}
\end{figure}

The following algorithm (also summarized in Figure \ref{fig:bi3+3_process}) 
\begin{enumerate}
    
    \item Enroll an mc of patients at dose $d$ in the main cohort. The default cohort size is 3. 
    \item Once the enrollment for mc is completed, start enrollment for bc by randomly allocating patients in bc to doses in ${\cal B}_d.$ Alternatively, one may restrict the enrollment of bc to dose $(d-1),$ depending on the specific situations of the trial. 
    \item After the DLT outcomes of patients at the mc are observed, obtain the dosing decision ${\cal S}_d \in \{E, S, D\}$ of dose $d$ based on the i3+3 design (Table \ref{tbl:i3+3_algorithm}), which is $E$, to escalate to dose $(d+1)$, $S$, to stay at dose $d$, or $D$ to de-escalate to dose $(d-1)$.  Do not execute the decision yet. Simply obtain and record it. 
    \item At the same time, follow the steps below and obtain the dosing decision ${\cal T}_k$ for dose $k$ lower than the current dose $d$. 
    \begin{enumerate}
        \item If one or more patients at dose $k$ are still being followed for DLT assessment without an outcome, apply the POD-i3 design  \citep{zhou2020review} to assess whether the enrollment must be suspended. If yes for any dose $k$, 
        suspend the trial enrollment at all doses and continue following the patients with pending DLT outcomes. 
        \item As time passes, eventually pending patients in the backfill cohort will complete follow up and POD-i3 will be able to generate a decision that is not to suspend the enrollment.  
        Then obtain a dosing decision ${\cal T}_k$ for dose $k$ using the POD-i3 design, which is $E$, to escalate to dose $k+1$, $D$, to de-escalate to dose $k-1$, or $S$, to stay at dose $k$.  Follow the steps below:
    \begin{enumerate}
        \item If ${\cal T}_k = D$ for at least one dose $k$, 
        find $k^* = \min\{k : {\cal T}_k = D\}$
        which corresponds to the lowest dose $k^*$ lower than the current dose $d$ 
        with  decision $D$. Enroll the next mc at dose $1\vee (k^* - 1)$.  When the starting  dose has a decision  $D$, i.e., $k^* = 1$, one could also consider inserting a lower dose instead.
        \item Otherwise,  enroll the next mc based on ${\cal S}_d$.
    \end{enumerate}
    \end{enumerate}
    \item Throughout the trial, if a new outcome is observed at any dose, apply the safety rule in the i3+3 design at all doses \citep{liu2020i3+},  which is detailed in $\S$\ref{sec:safety_rule}.  
    \item Repeat steps 1-5 until the number of patients in the main cohort reaches a maximum sample size or no doses are left due to safety rules.
\end{enumerate}

\noindent {\bf Remark 1} Since the backfill cohort can only be enrolled after the main cohort is enrolled, and once the main cohort completes follow up, another main cohort must be enrolled, the number of patients enrolled in the backfill cohort is random and  depends on the enrollment speed. If ${\cal B}_d$ is empty, there is no need to enroll backfill cohorts since no doses are allowed for backfill. 

\noindent {\bf Remark 2} In step 4, alternative approaches such as POD-i3+3 \citep{xu2022probability} may be applied as an alternative to POD-i3. The main feature of both designs is to consider a risk-speed tradeoff when it comes to making a decision based on incomplete DLT data.

\subsection{Determine the doses for backfill} 
\label{sec:update_d0}
In the Bi3+3 algorithm described above, 
when the mc is allocated to dose $d$, in principle all doses lower than $d$ are available for backfilling, i.e., ${\cal B}_d = \{k_0 = 1, 2, \cdots, d-1 \}.$ 
Due to the up-and-down nature of the decisions in Table \ref{tbl:i3+3_algorithm}, all doses lower than $d$ must have been tested and their safety has been established by the DLT data observed from the patients treated at these doses. Otherwise, it is impossible for the design to reach dose $d$ for mc if any lower dose exhibits excessive toxicity, in which case the dosing decision would not have been $E$, to escalate. 
Therefore, all the doses in ${\cal B}_d$ are deemed safe based on observed data. 
To this end, to determine ${\cal B}_d$, we consider efficacy data from patients and exclude low doses with insufficient efficacy. These doses are considered having little chance to induce beneficial efficacy response from cancer patients and therefore are not worth further exploration.

We follow the idea  in \citet{dehbi2021controlled}. We slightly abuse the notation $d$ hereinafter to either denote the current dose in the trial or a general index of any doses when needed. Occasionally, we also use $k$  or $i$   to index doses when $d$ is used to denote the current doses. Given context, these notations should be clear to readers. We start with   ${\cal B}_d = \{ k_0 = 1, 2, \cdots, d-1 \}$ assuming the current dose is $d$ for the mc.  
When a new efficacy outcome is observed from either mc or bc, we refresh the lowest dose $k_0$ in ${\cal B}_d$ based on the following procedure.  Recall the efficacy is assumed to be a binary outcome, e.g., objective response (OR) as in Response Evaluation Criteria in Solid Tumors (RECIST). 

To update $k_0$, we first define for a dose $k \in {\cal B}_d$,
  $$ q_{k+} = \frac{\sum_{i > k}^{D} n_i q_i}{\sum_{i > k}^{D} n_i}, $$ 
where $q_i$ is the true efficacy probability for dose $i$, and $n_i$ denotes the number of patients with available efficacy outcomes at dose $i$. By convention,  $0/0 = 0$. The quantity $q_{k+}$ represents the expected response rate at doses higher than dose $k$. 
Under a Bayesian model that will be presented later, we compute the posterior probability 
$\xi_k = \Pr\{q_{k+} > q_k \mid \mathcal{D}\}, \; k \in {\cal B}_d, $ where $\mathcal{D} = \{(v_d,n_d), d = 1,\cdots,D\}$, the observed efficacy data. 
If $\xi_k > \xi_0$, say $\xi_0 = 0.8,$ dose $k$ is deemed to be less efficacious than doses higher than $k$. The first dose $k_0$ in ${\cal B}_d$ is set to dose $(k+1)$, i.e.,  $k_0 = k+1$. 
Throughout the trial, posterior probabilities $\xi_k$ for $k \in {\cal B}_d$ are monitored continuously to determine $k_0$ and update ${\cal B}_d$. 

\subsection{Safety Rule} 
\label{sec:safety_rule}
Following the i3+3 design \citep{liu2020i3+},
two safety rules are added as ethical constraints to avoid excessive toxicity.  Both rules are applied in step 5 of the Bi3+3 design (Section \ref{sec:pat_allocation}).  
\begin{itemize}
    \item  \textbf{Rule 1 (Dose Exclusion)}: If the current dose is considered excessively toxic, i.e., $Pr\{p_d > p_T \mid \mathcal{H}\} > \eta$,   where $\mathcal{H} = \{(y_d,n_d), d = 1,\cdots,D\}$ and the threshold  $\eta$  is close to 1, say 0.95, the current and all higher doses are excluded and never be used again in the remainder of the trial.
    \item \textbf{Rule 2: Early Stop}: If the current dose is the lowest dose (the first dose) and is considered excessively toxic according to Rule 1,  stop the trial for safety. 
\end{itemize}

In safety rules 1 and 2, $ Pr\{p_d > p_T \mid \mathcal{H}\}$ is  under the posterior distribution 
$Beta(\alpha_0+y_d,\beta_0+n_d-y_d)$ which corresponds to an independent $Beta(\alpha_0, \beta_0)$ for $p_d$,   and $\alpha_0=\beta_0=1$ is used.  As a special case of the safety rule 1, if the decision for  the current dose $d$ is ``E" according to the decision rule in \S\ref{sec:pat_allocation}, and if the next higher dose $(d+1)$ level has been excluded due to Rule 1, the decision for the current dose is changed to  ``S" Stay for the current dose because there is no available dose to escalate to.

\section{MTD and OBD selection}
\label{sec:mtd_obd_selection}
\subsection{MTD selection}
\label{sec:mtd_selection}
After a prespecified maximum sample size of the trial is reached, all the DLT and efficacy data will be collected for all patients enrolled in the trial. We apply the same MTD selection procedure as in many recent designs based on the isotonic regression \citep{ji2010modified, guo2017bayesian, liu2020i3+}. In particular, an independent Beta(0.005, 0.005) prior is assumed for $p_d, \; d=1, \cdots, D.$
And the pooled adjacent violators algorithm (PAVA) \citep{robertson1988order} is applied to calculate the posterior mean toxicity probabilities for all doses, $\hat{p}_d$, $d=1,\cdots,D$, subject to the order constraints $\hat{p}_k > \hat{p}_d$ for $\forall k > d$. Among all tried doses, select as the estimated MTD $\widetilde{d}$ the dose with the smallest difference $\left| \hat{p}_d - p_T \right|$ and $\hat{p}_d \leq p_T + \epsilon_2$, i.e.,
$$\widetilde{d} = \mathop{\mathrm{argmin}}\limits_{\{d: n_d \neq 0, \hat{p}_d \leq p_T + \epsilon_2\}} \left| \hat{p}_d - p_T \right|.$$

\subsection{OBD selection}
\label{sec:obd_selection}
To select the  OBD,   we modify the change-point model in  \citet{dehbi2021controlled} and construct a four-parameter model for the dose-efficacy response. Specifically, recall that $q_d$ is the probability of efficacy at dose $d$. Then assume 
\begin{eqnarray*}
\text{logit}(q_d)=\beta_{0}+\beta_{1} \displaystyle{\left\{ x_d I(x_d \le h)+(\beta_2 +h)I(x_d >h)\right \}.}
\end{eqnarray*}
Here  $x_d \in \{1, \cdots, D\}$ denotes the $D$ discrete dose levels,  $\beta_0$ is the intercept, $\beta_1$ is the slope,  and  $h \in \{1, \cdots, D\}$ is the change point of the model  after which   efficacy  is assumed to plateau.  Under this model, when the dose level is less than or equal to $h$, by assuming $\beta_1 > 0$, there is a positive monotonic relationship between dose level and efficacy response rate; when the dose level is higher than $h$, efficacy response rate no longer increases with dose level and stays  as constant  $\beta_0 + \beta_1(\beta_2 +h)$ at the logit scale.   Quantity $\beta_1 
 \beta_2$ represents the increment in comparison to the  response rate $\beta_0 + \beta_1 h$ at the change point $h$.  This increment is needed to describe the response rate of the dose immediately after the change point $h$ due to discrete dosing.  

The prior  of $\beta_1$ is set as log-normal distribution with mean 0 and a large variance. Parameter $\beta_2$ describes a jump after the change point in response rate, and its prior distribution follows another log-normal. We use  a normal prior with mean  -2 for $\beta_0$ so that the prior probability of response when there is no treatment is small. As for parameter $h$, we set a discrete prior where $h$ takes the value of each dose level from the starting dose 1 to the highest dose $D$  with a probability. Specifically, we 
set the prior probability to be a small value $e$, say $e=0.05,$ for any dose with no enrolled patients, and divide the remaining prior probability  evenly across doses with enrolled patients.   We use $D' \le D$ to denote the number of doses with enrolled patients.  The  prior distributions are given as follows.
\begin{eqnarray*}
    \beta_0  \sim  N(-2,10), 
    \beta_1  \sim  LN(0,10), 
    \beta_2  \sim  LN(0,10), \mbox{ and }\\
    h  \sim  Cat(r_{1}, \cdots, r_{D}) \mbox{ where } r_d = e * I(n_d =0) + \frac{1-(D-D')*e}{D'}I(n_d >0),
\end{eqnarray*}
  where $Cat(r_{1}, \cdots, r_{D})$ is a discrete distribution taking values in $\{1, \cdots, D\}$ with corresponding probabilities $(r_{1}, \cdots, r_{D})$, respectively.    Note $r_d$ is simply the mathematical expression that gives us the aforementioned prior probability for $h$.

Given available efficacy data $\mathcal{D} = \{(v_d,n_d), d = 1,\cdots,D\}$, the posterior probability $\phi_d = \Pr\{h = d \mid \mathcal{D}\}$, $d = 1,\cdots,D$, can be calculated and  used to select OBD. The change point  $h^*$ is estimated as
$$h^* = D' \wedge ( \mathop{\mathrm{argmax}}\limits_{d} \phi_d ).$$
Considering both toxicity and efficacy effects of these doses, we finally select the OBD $d^*$ based on the selected MTD $\widetilde{d}$ and the estimated change point $h^*$ as
$$d^* = \widetilde{d} \wedge ( h^* + 1 ).$$


\section{Simulation}
\label{sec:simulation}
  We conduct simulated trials to assess the performance of the Bi3+3 design. We generate patient toxicity and efficacy data based on a set of scenarios. 
For each scenario, we run 1,000 simulated trials.    To  mimic  real-world situations, we assume the toxicity and efficacy outcomes of any patients are   not immediately observed. Instead, we assume the time to DLT follows a uniform distribution ranging from 0 to the maximum DLT follow-up time if a DLT occurs for the patient;  otherwise we assume DLT is censored.  The efficacy outcomes are observed 90 days after patients' enrollment. Also, the arrival time of patients is assumed to follow an exponential distribution with a mean of 10 days, which means, on average, every 10 days a new patient is eligible for enrollment of the trial, and hence the trial would enroll about three patients per month.

In $\S$\ref{sec:comparison_bcrm},we specify  several scenarios with different target toxicity probabilities  and different numbers of doses, and compare Bi3+3 with the Backfill CRM design \citep{dehbi2021controlled}. In $\S$\ref{sec:com_mTPI2} we compare Bi3+3 with the mTPI-2 design \citep{guo2017bayesian} for trial duration.  

\subsection{Comparison with Backfill CRM}
\label{sec:comparison_bcrm}
We first compare Bi3+3 with the Backfill CRM design in \citet{dehbi2021controlled}.    
We briefly summarize the Backfill CRM design below, which provides instructions for allocating patients in the main cohort and backfill cohort. 
\begin{itemize}
    \item Main cohort allocation
    \begin{enumerate}
        \item A one-parameter CRM model \citep{o1990continual} is used for dose assignment and MTD selection, only using data in the main cohort.
    \end{enumerate}
    \item Backfill cohort allocation
    \begin{enumerate}
        \item Starting with the second cohort of patients in the main cohort, a backfill cohort with three patients will be enrolled after the main cohort enrollment is complete.
        \item The patients are randomly assigned to backfill doses  with equal probability. 
        \item Similar to the rules in $\S$\ref{sec:update_d0}, doses with low efficacy response rates are removed from backfill doses.   
    \end{enumerate}
    \item OBD selection
    \begin{enumerate}
        \item After the trial ends, model selection is performed to select a monotonic logistic regression {or} a change-point logistic regression.
        \item The OBD is decided based on the selected model. If the change-point model is selected, doses are divided to two parts, monotone and plateau parts. The dose on the plateau and closest to the change point is selected as the OBD.
    \end{enumerate}
\end{itemize}

\begin{figure}[!htbp]
    \centering
    \includegraphics[width=1.0\textwidth]{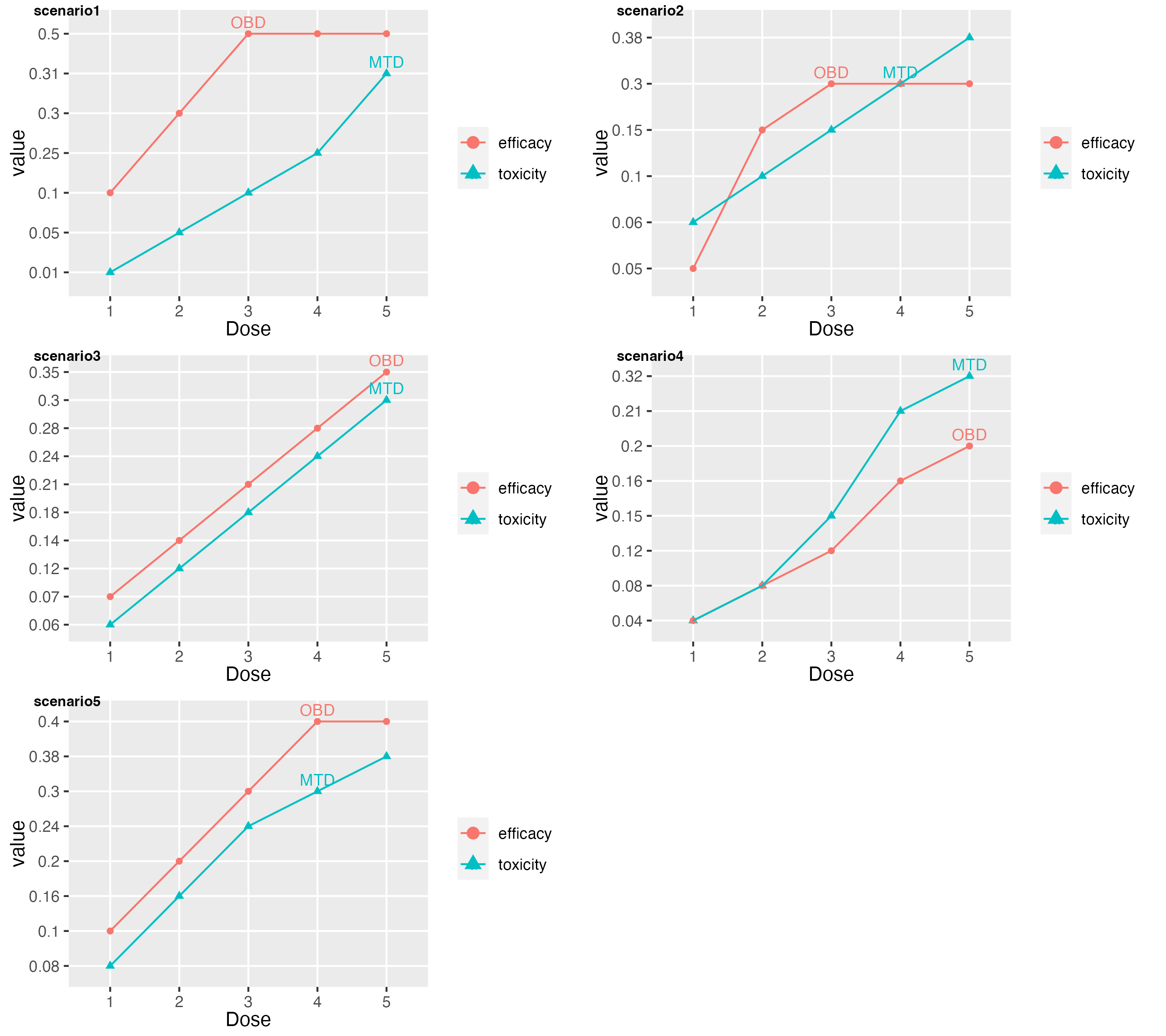}
    \caption{  The five simulation scenarios with target toxicity probability of 0.3. The (MTD, OBD) are (5,3), (4,3), (5,5), (5,5), and (4,4) for scenarios 1-5, respectively.  } 
    \label{fig:sc_vs_crm}
\end{figure}

We implement the Backfill CRM design and benchmark our implementation against the simulation results in  \citet{dehbi2021controlled}. 
Our results, shown in Appendix \ref{appendix:reproduction}, are very close to those presented in \cite{dehbi2021controlled}, thus  assuring   our implementation. We then compare the Bi3+3 design with Backfill CRM on a variety of scenarios. See Figure \ref{fig:sc_vs_crm} for the five scenarios used in our simulation. 
In the subsequent simulation,    the target probability of toxicity for MTD  is set to  $p_T =0.3,$   and the equivalence interval for Bi3+3 is  $EI=[0.25,0.35].$  As the backfill strategy is different in Bi3+3 design and Backfill CRM design,   we try to match the overall sample size (main cohorts + backfill cohorts) for both designs for fair comparison.  For the Bi3+3 design,  30 patients are enrolled in the main cohorts which result in an average sample size of 42.8, with the additional 12.8 patients coming from the backfill cohorts.  As for the Backfill CRM design, we set 24 as the main cohort sample size which produces an average of 45.0 patients for the entire trial.  We use the skeleton in \cite{lee2011calibration} for Backfill CRM, a popular choice in practice.    
The operating characteristics of both designs are  presented in Table \ref{tbl:oc_vs_crm}. 

\begin{table}[!htbp]
\scriptsize
\begin{center}
\caption{Comparison of Bi3+3 and Backfill CRM in five scenarios. The target toxicity probability is 0.3 and EI =[0.25,0.35].} 
\label{tbl:oc_vs_crm}
\begin{tabular}{@{}cccccc|ccccc@{}}
\toprule
 & \multicolumn{5}{c|}{Bi3+3}  & \multicolumn{5}{c}{Backfill CRM}\\ \midrule
 Dose level &  1&2  &3  &4  &5 &  1&2  &3  &4  &5 \\ \midrule
\textbf{Scenario 1}&  &  &  &  & &  &  &  &  &  \\
Efficacy probability& 0.1 & 0.3 &0.5  &0.5  &0.5 & 0.1 & 0.3 &0.5  &0.5  &0.5 \\
Toxicity probability& 0.01 &0.05  & 0.1  &0.25  & 0.31 & 0.01 &0.05  & 0.1  &0.25  & 0.31\\
\% of OBD selection& 0\%  & 17.4\% &\textbf{34.3\%}  &23.7\%  &24.6\%  & 0\% &0.5\%  &\textbf{40.1\%} &43.4\%  &16.0\% \\
\% of MTD selection& 0\% & 0.5\% &  20.5\% & 37.3\% & \textbf{41.7\%} &0\% &0.2\%  &5.0\%  &35.0\%  &\textbf{59.8\%} \\
Efficacy estimation& 0.22 &0.32  &0.43  &0.48  &0.52 &0.24  &0.32  & 0.41 & 0.47 &0.51 \\ 
Number of patients enrolled & 7.1 & 7.4 & \textbf{10.3} & 10.3 & \textbf{8.1}&8.4  &9.3  & \textbf{10.3} &9.7  &\textbf{7.3}\\
Number of patients backfill &  4.0 &3.6 &3.6 &1.9 &0& 5.3 & 6.2 & 6.2 & 3.3 &0\\
\textbf{Scenario 2}&  &  &  &  &  &  &  &  &  &\\
Efficacy probability& 0.05 &0.15  & 0.3 &0.3  &0.3& 0.05 &0.15  & 0.3 &0.3  &0.3 \\
Toxicity probability&0.06 &0.1  & 0.15  &0.3  & 0.38& 0.06 &0.1  & 0.15  &0.3  & 0.38 \\
\% of OBD selection& 0.5\%  & 13.9\% &\textbf{39.2\%}  & 31.3\% &15.1\%  & 0\% &5.2\%  &\textbf{49.9\%}  &37.7\%  &7.2\%\\
\% of MTD selection& 0.5\% & 3.7\% & 33.7\% & \textbf{40.8\%} & 21.3\%& 0\% & 2.5\% & 22.2\% & \textbf{50.2\%} & 25.1\%\\
Efficacy estimation& 0.13 &0.18  &0.24  &0.29  &0.34 &0.14&0.18&0.23&0.28&0.32 \\ 
Number of patients enrolled& 8.6 & 8.7 & \textbf{10.9} & \textbf{9.3} & 4.7& 11.2 &10.3  &\textbf{10.7} &\textbf{8.4}  &4.4\\
Number of patients backfill & 4.8  &3.7 &2.7 &1.0 &0& 7.3 &6.5  &5.2 &2.0  &0\\
\textbf{Scenario 3}&  & &  &  & &  &  &  &  & \\
Efficacy probability& 0.07 &0.14  &0.21  &0.28  &0.35& 0.07 &0.14  &0.21  &0.28  &0.35 \\
Toxicity probability& 0.06 &0.12  &0.18  &0.24  &0.3 &0.06 &0.12  &0.18  &0.24  &0.3\\
\% of OBD selection& 0.8\% & 12.3\% & 26.3\% & 30.1\% & \textbf{30.5\%} &  0.2\% &8.3\%  &39.4\%  & 35.2\% & \textbf{16.9\%}\\
\% of MTD selection& 0.8\% & 6.4\% & 23.4\% & 33.3\% & \textbf{36.1\%} & 0.2\% &3.4\%  &21.0\%  &42.7\% & \textbf{32.7\%}\\
Efficacy estimation& 0.12 & 0.16 &0.21  & 0.26 &0.31 &0.13 &0.16& 0.20 &0.25  &0.29 \\ 
Number of patients enrolled & 9.5 & 9.7 & 10.1 & 7.8 & \textbf{6.0}&11.8  &10.5 &10.0 &7.8  &\textbf{4.9}\\
Number of patients backfill & 5.5  &3.8 &2.6 &1.2 &0& 7.9 & 6.4 &4.5  &2.2  &0\\
\textbf{Scenario 4}&  &  &  &  &&  &  &  &  &  \\
Efficacy probability& 0.04 &0.08  &0.12  &0.16  &0.2& 0.04 &0.08  &0.12  &0.16  &0.2 \\
Toxicity probability& 0.04 &0.08  &0.15  &0.21  &0.32& 0.04 &0.08  &0.15  &0.21  &0.32\\
\% of OBD selection& 0.1\% & 6.0\% &15.6\%  & 38.2\% & \textbf{40.1\%}&  0.1\%  &2.6\%  &43.9\%  &31.7\%  &\textbf{21.7\%} \\
\% of MTD selection& 0.1\% & 3.4\% & 14.3\% & 39.9\%  & \textbf{42.3\%}&  0.1\%  & 0.5\% &10.2\%  &41.0\%  &\textbf{48.2\%} \\
Efficacy estimation&  0.08& 0.10 & 0.12  &0.15  &0.19 & 0.09 &0.10  &0.12  &0.14  &0.17 \\ 
Number of patients enrolled & 9.2 & 8.7 & 9.8 & 9.0 & \textbf{7.7}& 10.7 & 9.9 & 9.7& 8.4 &\textbf{6.4}\\
Number of patients backfill & 5.7 &4.0 &3.0 &1.6 &0& 7.2 &6.4  &4.9  &2.5  &0\\
\textbf{Scenario 5}&  &  &  &  & &  &  &  &  & \\
Efficacy probability& 0.1 &0.2  &0.3  &0.4  &0.4& 0.1 &0.2  &0.3  &0.4  &0.4 \\
Toxicity probability& 0.08 &0.16  & 0.24  &0.3  & 0.38& 0.08 &0.16  & 0.24  &0.3  & 0.38 \\
\% of OBD selection& 3.5\% & 27.7\% & 31.3\% & \textbf{25.3\%} & 12.1\% & 0.1\%   &16.7\%  &49.6\% &\textbf{29.5\%}&4.1\%\\
\% of MTD selection& 3.5\% & 19.7\% & 31.9\% & \textbf{29.2\%} & 15.6\% &  0.1\%  & 9.9\% &41.1\%  & \textbf{39.4\%} &9.5\%\\
Efficacy estimation& 0.16 & 0.21 &0.28  &0.36  &0.40 &0.17  &0.21  & 0.27 & 0.33 &0.37 \\ 
Number of patients enrolled & 10.3 & 11.4 & 10.2 & \textbf{6.2} & 3.2&13.1  &11.9  &  10.6& \textbf{6.4} &2.9\\ 
Number of patients backfill &5.3 &3.3 &1.8 &0.7 &0&8.7  &6.8  &4.0  &1.5  &0\\\bottomrule
\end{tabular}
\end{center}
\end{table}


For scenarios 1 and 2, the true MTD is the fifth dose and fourth dose, respectively. The efficacy response rates of doses reach a plateau at the third dose, which is the true OBD. In scenarios 1 and 2, Bi3+3 selects the true MTD in 41.7\% and 40.8\% of the simulated trials, highest among all the doses, and it selects the true OBD, dose level three, 34.3\% and 39.2\% of the times. The Backfill CRM design has slightly higher percentages in OBD selection in these two scenarios although it also tends to select the next  higher dose with a high frequency. 
On average, Bi3+3 assigns 10.3 patients to dose levels three and four in scenario 1, respectively. Compared with Backfill CRM, the total sample size of Bi3+3 is 2.2 patients smaller, although there is not much difference in the patient  allocation to the OBDs. In scenarios 3 and 4 both toxicity and efficacy increase with dose level and the MTD and OBD are the same, which is dose level five. The Bi3+3 model identifies dose five as the MTD, with frequencies of 36.1\% and 42.3\%, and as the Optimal Biological Dose (OBD), with frequencies of 30.5\% and 40.1\% in scenarios 3 and 4, respectively. In contrast, the Backfill CRM design selects dose three as OBD with probabilities 39.4\% and 43.9\% in the two scenarios, which is less desirable.    
For patient allocation, Bi3+3 allocates on average 6.0 and 7.7 patients to dose five compared to 4.9  and 6.4 by  Backfill CRM  in scenarios 3 and 4, respectively. In the last scenario 5, efficacy plateaus at the true MTD, indicating that the true MTD is the same as the true OBD, which is dose four. However, dose three has a toxicity rate close to the MTD, which makes it difficult for both designs to differentiate the two doses. This can be seen in the selection percentages of the MTD and OBD for both designs. 
Across all the scenarios, Backfill CRM is slightly more conservative than Bi3+3 as it allocates more patients to lower doses. However, it tends to select doses in the middle. This could be desirable if the true OBD is near the middle like in scenarios 1 and 2.  In contrast, in cases where the true OBD is at a high dose, like scenarios 3 and 4, Backfill CRM may have less chance finding them. In summary, both designs exhibit desirable features in these five scenarios.

Apart from these features, we also calculate the mean of efficacy using the regression model in each scenario. The results are reported in Table \ref{tbl:oc_vs_crm} corresponding to the line ``Efficacy estimation". The estimated efficacy response rates in general do not appear to deviate much from the true efficacy probabilities. 

\subsection{Comparison with   the    mTPI-2 design given $p_T=0.3$}
\label{sec:com_mTPI2}
Next, we apply mTPI-2 \citep{guo2017bayesian} to illustrate some features of Bi3+3 as mTPI-2 is a design that does not allow patient backfill. 
For fair comparison, we first enroll 30 patients for mTPI-2, select an MTD, and allow thirteen more patients to be enrolled at the selected MTD (to adjust for the fact that Bi3+3 would enroll more patients in the backfill cohorts). After the additional thirteen  patients are enrolled and their DLT data are observed, we re-select the MTD based on the additional data. We report the frequency of the re-selected MTD in the simulation results for mTPI-2. 

The results comparing Bi3+3 and mTPI-2 are presented in Table \ref{tbl:oc_vs_mtpi2_pt0.3}. Bi3+3 has an advantage in many scenarios as it allows patients to be backfilled. For example, it does not increase trial duration while allowing more patients to be assigned to lower doses, resulting in better OBD selection. See Table \ref{tbl:oc_vs_mtpi2_pt0.3} for a summary.    

As a sensitive analysis, we vary the target toxicity probability from $p_T=0.3$ to $p_T=0.25$ and change the EI from [0.25,0.35] to [0.2,0.3] (Table \ref{tbl:oc_vs_mtpi2_pt0.25}).    
We also construct four more scenarios to assess the trial duration and patients enrollment. Bi3+3 shows desirable results as expected, thanks to the ability to backfill patients and time-to-event modeling (Table \ref{tbl:duration_vs_mtpi2}).    

\section{Discussion}
\label{sec:discussion}
The backfill strategy in phase I trials has been routinely applied to allow accumulation of information at low but potentially efficacious doses, despite lacking formal statistical designs. Here, we propose a statistical framework based on model-based inference to allow patient backfilling. The main contributions are to apply POD for patients with pending outcomes and to model potentially plateau dose-efficacy response for OBD selection. We show that the proposed Bi3+3 design is able to put more patients at lower doses without increasing the trial duration. 

For the selection of OBD, we develop a four-parameter regression model that assumes a change-point dose-response relationship. More complex models may be considered, such as model average over possible dose-efficacy models. These models may require a larger sample size as they involve more parameters. In general, tradeoff between statistical principles on using optimal models and simplicity and interpretability of results must be balanced. Also, in recent oncology drug development, occasionally a new drug may manifest non-monotone dose-efficacy relationship in which the efficacy may initial increase but later decrease with dose. Therefore, a second-order dose-response model might be needed to address such relationship. 

 In Bi3+3, for simplicity we apply the POD-i3 design at a backfill dose. The POD-i3 design only uses the data from that dose to generate a dosing decision.  Alternatively, one could combine the POD framework with designs like CRM (i.e., POD-CRM) so that data from all the doses are used for statistical inference and decision making at each backfill dose. This will be left for future work. 

We adopt the i3+3 design for the main cohort dose escalation. This is optional. Any sensible design may work well under the proposed framework. In addition,  no constraints are imposed on the number of patients enrolled at a single dose. In practice,  for the purpose of controlling resources, a max number of patients at a dose can be specified,  such as 12 patients.

\clearpage

\begin{table}[!htbp]
\tiny
\begin{center}
\caption{Comparison of Bi3+3 and mTPI-2 in five scenarios. The target toxicity probability is 0.3, EI =[0.25,0.35], sample size is 42.8 for Bi3+3 and 43.0 for mTPI-2.}
\label{tbl:oc_vs_mtpi2_pt0.3}
\begin{tabular}{@{}cccccc|ccccc@{}}
\toprule
 & \multicolumn{5}{c|}{Bi3+3}  & \multicolumn{5}{c}{mTPI-2}\\ \midrule
 Dose level &  1&2  &3  &4  &5 &  1&2  &3  &4  &5 \\ \midrule
\textbf{Scenario 1}&  &  &  &  & &  &  &  &  &  \\
Efficacy probability& 0.1 & 0.3 &0.5  &0.5  &0.5  &  &  &  &  &\\
Toxicity probability& 0.01 &0.05  & 0.1  &0.25  & 0.31 & 0.01 &0.05  & 0.1  &0.25  & 0.31\\
\% of OBD selection& 0\%  & 17.4\% &\textbf{34.3\%}  &23.7\%  &24.6\%   &  &  &  &  & \\
\% of MTD selection&  0\% & 0.5\% &  20.5\% & 37.3\% & \textbf{41.7\%}&0\% &0.4\%  &20.5\%  &42.4\%  &\textbf{36.7\%} \\
Efficacy estimation& 0.22 &0.32  &0.43  &0.48  &0.52 &  &  &  &  & \\ 
Number of patients enrolled & 7.1 & 7.4 & \textbf{10.3} & 10.3 & \textbf{8.1}&3.1  &3.8  & \textbf{8.7} &13.4  &\textbf{14.0}\\
Number of patients backfill &   4.0 &3.6 &3.6 &1.9 &0&  &  &  &  &\\
Trial Duration& \multicolumn{5}{c}{494}  & \multicolumn{5}{c}{619}\\
\textbf{Scenario 2}&  &  &  &  &  &  &  &  &  &\\
Efficacy probability& 0.05 &0.15  & 0.3 &0.3  &0.3&  &  &  &  &  \\
Toxicity probability&0.06 &0.1  & 0.15  &0.3  & 0.38& 0.06 &0.1  & 0.15  &0.3  & 0.38 \\
\% of OBD selection& 0.5\%  & 13.9\% &\textbf{39.2\%}  & 31.3\% &15.1\%  &  &  &  &  &\\
\% of MTD selection& 0.5\% & 3.7\% & 33.7\% & \textbf{40.8\%} & 21.3\%& 0.1\% & 3.5\% & 39.4\% & \textbf{41.7\%} & 15.3\%\\
Efficacy estimation& 0.13 &0.18  &0.24  &0.29  &0.34 &  &  &  &  & \\ 
Number of patients enrolled&  8.6 & 8.7 & \textbf{10.9} & \textbf{9.3} & 4.7& 3.8 &5.4  &\textbf{12.1} &\textbf{13.7}  &8.0\\
Number of patients backfill & 4.8  &3.7 &2.7 &1.0 &0&  &  &  &  &\\
Trial Duration& \multicolumn{5}{c}{490}  & \multicolumn{5}{c}{615}\\
\textbf{Scenario 3}&  & &  &  & &  &  &  &  & \\
Efficacy probability& 0.07 &0.14  &0.21  &0.28  &0.35&  &  &  &  &  \\
Toxicity probability& 0.06 &0.12  &0.18  &0.24  &0.3 &0.06 &0.12  &0.18  &0.24  &0.3\\
\% of OBD selection& 0.8\% & 12.3\% & 26.3\% & 30.1\% & \textbf{30.5\%}  &  &  &  &  &\\
\% of MTD selection& 0.8\% & 6.4\% & 23.4\% & 33.3\% & \textbf{36.1\%} &0.2\% &6.6\%  &25.5\%  &35.9\% & \textbf{31.8\%}\\
Efficacy estimation& 0.12 & 0.16 &0.21  & 0.26 &0.31 &  &  &  &  & \\ 
Number of patients enrolled & 9.5 & 9.7 & 10.1 & 7.8 & \textbf{6.0}&4.0  &6.7 &10.4 &10.7  &\textbf{11.1}\\
Number of patients backfill & 5.5  &3.8 &2.6 &1.2 &0&  &  &  &  &\\
Trial Duration& \multicolumn{5}{c}{493}  & \multicolumn{5}{c}{619}\\
\textbf{Scenario 4}&  &  &  &  &&  &  &  &  &  \\
Efficacy probability& 0.04 &0.08  &0.12  &0.16  &0.2&  &  &  &  &  \\
Toxicity probability& 0.04 &0.08  &0.15  &0.21  &0.32& 0.04 &0.08  &0.15  &0.21  &0.32\\
\% of OBD selection& 0.1\% & 6.0\% &15.6\%  & 38.2\% & \textbf{40.1\%}&  &  &  &  & \\
\% of MTD selection& 0.1\% & 3.4\% & 14.3\% & 39.9\%  & \textbf{42.3\%}& 0\% & 2.9\% &17.0\%  &45.7\%  &\textbf{34.4\%} \\
Efficacy estimation&  0.08& 0.10 & 0.12  &0.15  &0.19 &  &  &  &  & \\ 
Number of patients enrolled &  9.2 & 8.7 & 9.8 & 9.0 & \textbf{7.7}& 3.5 & 4.9 & 8.6& 12.5 &\textbf{13.4}\\
Number of patients backfill &  5.7 &4.0 &3.0 &1.6 &0&  &  &  &  &\\
Trial Duration& \multicolumn{5}{c}{495}  & \multicolumn{5}{c}{621}\\
\textbf{Scenario 5}&  &  &  &  & &  &  &  &  & \\
Efficacy probability& 0.1 &0.2  &0.3  &0.4  &0.4 &  &  &  &  & \\
Toxicity probability& 0.08 &0.16  & 0.24  &0.3  & 0.38& 0.08 &0.16  & 0.24  &0.3  & 0.38 \\
\% of OBD selection&  3.5\% & 27.7\% & 31.3\% & \textbf{25.3\%} & 12.1\% &  &  &  &  &\\
\% of MTD selection& 3.5\% & 19.7\% & 31.9\% & \textbf{29.2\%} & 15.6\% &2.6\% & 21.8\% &34.7\%  & \textbf{30.2\%} &10.6\%\\
Efficacy estimation&0.16 & 0.21 &0.28  &0.36  &0.40  &  &  &  &  & \\ 
Number of patients enrolled & 10.3 & 11.4 & 10.2 & \textbf{6.2} & 3.2&5.1  &10.2  &  12.6& \textbf{9.4}\textbf{} &5.6\\ 
Number of patients backfill &5.3 &3.3 &1.8 &0.7 &0&  &  &  &  &\\
Trial Duration& \multicolumn{5}{c}{489}  & \multicolumn{5}{c}{613}\\\bottomrule
\end{tabular}
\end{center}
\end{table}


\begin{table}[!htbp]
\scriptsize
\begin{center}
\caption{Comparison of trial duration and patients' enrollment of Bi3+3 and mTPI-2}
\label{tbl:duration_vs_mtpi2}
\begin{tabular}{cccccccccc}
\hline
\multirow{2}{*}{} & \multicolumn{5}{c}{\textbf{Dose Toxicity}} & \multirow{2}{*}{\textbf{\makecell{Bi3+3\\Trial Duration}}} & \multirow{2}{*}{\textbf{\makecell{mTPI-2\\Trial Duration}}} & \multirow{2}{*}{\textbf{\makecell{Bi3+3\\Patients enrolled}}} & \multirow{2}{*}{\textbf{\makecell{mTPI-2\\Patients enrolled}}} \\
                  & 1   &  2  &3    &4   & 5  &                   &                   &                   &                   \\
\textbf{Scenario 1}& 0.15&0.3& 0.45 &0.6& 0.75 &476 & 588&37.0&42.5               \\ 
\textbf{Scenario 2}&0.06&0.12&0.18&0.24& 0.44& 491&613 &42.7&43                   \\
\textbf{Scenario 3}&0.05&0.1&0.15&0.2&0.25&495&625&43.9&43                   \\
\textbf{Scenario 4}&0.27&0.37&0.47&0.57&0.67&435 &541&31.4&38.3                   \\
\textbf{Total Mean}&    &    &    &   &   &474.2&591.9&38.7&41.7   \\ \hline
\end{tabular}
\end{center}
\end{table}

\clearpage
\newpage

\bibliographystyle{apalike}
\bibliography{bi3+3}

\begin{thebibliography}{}

\bibitem[Barnett et~al., 2023]{Barnett2023}
Barnett, H., Boix, O., Kontos, D., and Jaki, T. (2023).
\newblock Backfilling cohorts in phase i dose-escalation studies.
\newblock {\em Clinical trials}, 20(3):261--268.

\bibitem[Cheung and Chappell, 2000]{cheung2000sequential}
Cheung, Y.~K. and Chappell, R. (2000).
\newblock {Sequential designs for phase I clinical trials with late-onset
  toxicities}.
\newblock {\em Biometrics}, 56(4):1177--1182.

\bibitem[Dehbi et~al., 2021]{dehbi2021controlled}
Dehbi, H.-M., O’Quigley, J., and Iasonos, A. (2021).
\newblock {Controlled backfill in oncology dose-finding trials}.
\newblock {\em Contemporary Clinical Trials}, 111:106605.

\bibitem[{FDA}, 2023]{Optimus}
{FDA} (2023).
\newblock {Project Optimus}.
\newblock
  \url{https://www.fda.gov/about-fda/oncology-center-excellence/project-optimus}.

\bibitem[Guo et~al., 2017]{guo2017bayesian}
Guo, W., Wang, S.-J., Yang, S., Lynn, H., and Ji, Y. (2017).
\newblock {A Bayesian interval dose-finding design addressingOckham's razor:
  mTPI-2}.
\newblock {\em Contemporary clinical trials}, 58:23--33.

\bibitem[Ivanova et~al., 2007]{Ivaccd2007}
Ivanova, A., Flournoy, N., and Chung, Y. (2007).
\newblock Cumulative cohort design for dose-finding.
\newblock {\em Journal of Statistical Planning and Inference},
  137(7):2316--2327.

\bibitem[Ji et~al., 2010]{ji2010modified}
Ji, Y., Liu, P., Li, Y., and Nebiyou~Bekele, B. (2010).
\newblock {A modified toxicity probability interval method for dose-finding
  trials}.
\newblock {\em Clinical trials}, 7(6):653--663.

\bibitem[Lee and Cheung, 2011]{lee2011calibration}
Lee, S. and Cheung, Y. (2011).
\newblock Calibration of prior variance in the bayesian continual reassessment
  method.
\newblock {\em Statistics in medicine}, 30:2081--9.

\bibitem[Liu et~al., 2020]{liu2020i3+}
Liu, M., Wang, S.-J., and Ji, Y. (2020).
\newblock {The i3+ 3 design for phase I clinical trials}.
\newblock {\em Journal of biopharmaceutical statistics}, 30(2):294--304.

\bibitem[Liu and Yuan, 2015]{yuan2015boin}
Liu, S. and Yuan, Y. (2015).
\newblock Bayesian optimal interval designs for phase i clinical trials.
\newblock {\em Journal of the Royal Statistical Society. Series C (Applied
  Statistics)}, 64(3):507--523.

\bibitem[Manasanch et~al., 2020]{manasanch2020phase}
Manasanch, E.~E., Shah, J.~J., Lee, H.~C., Weber, D.~M., Thomas, S.~K., Amini,
  B., Olsem, J., Crumpton, B., Morphey, A., Berkova, Z., et~al. (2020).
\newblock {Phase I/Ib study of carfilzomib and panobinostat with or without
  dexamethasone in patients with relapsed/refractory multiple myeloma}.
\newblock {\em Haematologica}, 105(5):e242.

\bibitem[O'Quigley et~al., 1990]{o1990continual}
O'Quigley, J., Pepe, M., and Fisher, L. (1990).
\newblock {Continual reassessment method: a practical design for phase 1
  clinical trials in cancer}.
\newblock {\em Biometrics}, pages 33--48.

\bibitem[Robertson, 1988]{robertson1988order}
Robertson, T. (1988).
\newblock {Order restricted statistical inference}.
\newblock Technical report.

\bibitem[Shah et~al., 2021]{shah2021drug}
Shah, M., Rahman, A., Theoret, M.~R., and Pazdur, R. (2021).
\newblock The drug-dosing conundrum in oncology-when less is more.
\newblock {\em The New England journal of medicine}, 385(16):1445--1447.

\bibitem[Xu and Lin, 2022]{xu2022probability}
Xu, Z. and Lin, X. (2022).
\newblock Probability-of-decision interval 3+ 3 (pod-i3+ 3) design for phase i
  dose finding trials with late-onset toxicity.
\newblock {\em Statistical Methods in Medical Research}, 31(3):534--548.

\bibitem[Zhou et~al., 2020]{zhou2020pod}
Zhou, T., Guo, W., and Ji, Y. (2020).
\newblock {Pod-tpi: Probability-of-decision toxicity probability interval
  design to accelerate phase I trials}.
\newblock {\em Statistics in Biosciences}, 12(2):124--145.

\bibitem[Zhou and Ji, 2020]{zhou2020review}
Zhou, T. and Ji, Y. (2020).
\newblock Statistical frameworks for time-to-event dose-finding designs: A
  review.
\newblock {\em arXiv preprint arXiv:2006.11676}.

\end{thebibliography}

\newpage

\begin{appendices}
\appendixpage
\setcounter{table}{0}
\renewcommand{\thetable}{A.\arabic{table}}
\setcounter{figure}{0}
\renewcommand{\thefigure}{A.\arabic{figure}}
\section{The POD-i3 design} \label{appendix:pod_method}
After the outcomes of patients in the main cohort (mc) are observed, some pending patients at lower doses are still being followed. Generally, trial designs would make a dose decision immediately after the observation of outcomes at the main cohort and explore more doses as soon as possible. The issue is how the partial information from the pending patients can be incorporated for the dosing decision. We follow the POD framework \citep{zhou2020pod} to analyze and utilize information of pending outcomes. Specifically, we develop a POD-i3 design as suggested by the authors. 

Unlike the TITE-CRM design \citep{cheung2000sequential}, the POD framework considers probability of making a dosing decision ($E$, $S$, or $D$) where the randomness is induced by the uncertainty in the pending DLT outcomes. Let $\mathcal{A}_d$ denote the decision function of a complete-data design for the current dose $d$, such as the i3+3 design, which requires the complete DLT evaluation of all the enrolled patients. Given the target toxicity probability $p_T$ and EI$=[p_T - \epsilon_1, p_T + \epsilon_2]$, the decision of dose $d$, $\mathcal{A}_d$, only depends on $n_d$ and $y_d$, i.e., a deterministic function of $n_d$ and $y_d$, $\mathcal{A}_d = \mathcal{A}_d[y_d,n_d] \in \{-1, 0, 1\}$, where -1, 0, and 1 correspond to the decisions of de-escalating to the next lower dose (-1), stay at the current dose (0), and escalating to the next higher dose (1), respectively. Then, we use $y_{k,\obs}$ to denote the  observed toxicity outcomes at a backfill dose $k$ ($k < d$) and denote the unobserved DLT outcomes as  $Y_{k,\mis}$. Thus, letting $a$ takes values in $\{-1,0,1\}$, the decision function in the presence of pending outcomes at dose $k$ can be expressed as 
$A_k=\mathcal{A}_k[(y_{k,\obs},Y_{k,\mis}),n_k]$. As $Y_{k,\mis}$ is a random variable, the POD method calculates the posterior probability $\gamma_{k,a}\equiv Pr(A_k=a \mid \mathcal{H})$ as 
$$\gamma_{k,a} =\displaystyle\sum\limits_{y_{k,\mis}:{A}_k=a} \Pr(Y_{k,\mis}=y_{k,\mis} \mid \mathcal{H}), \quad k =  k_0,\cdots, d-1, \;\; \mbox{ and } a \in \{-1, 0, 1\}.$$
Here, $\mathcal{H}$ denotes the observed data of all enrolled patients, including the dose assigned to each patient, whether or not each patient experiences DLT within the follow-up window, and the follow-up time of each patient. The posterior probability $\gamma_{k,a}$ is the POD for decision $a$ at dose $k$.

Let $A^*_k = \mathop{\mathrm{argmax}}\limits_{a} \gamma_{k,a}$ denotes the decision at dose $k$ with the highest POD. If multiple decisions tie for the highest POD, we choose the more conservative one. To ensure the safety of the design, we apply a suspension rule similar to \cite{zhou2020pod}. If $A^*_k = 0$, i.e.,  stay at dose $k$, we recommend to suspend the trial if the posterior probability of de-escalation $\gamma_{k,-1} > \pi_D$ for a pre-specified threshold $\pi_D$. In the original paper by \cite{zhou2020pod}, there is another suspension rule for dose escalation. However, since we are working with backfill doses which have already shown initial safety, we decide not to enforce that rule. See more detail in \cite{zhou2020pod}.

\section{Benchmark Results for the Backfill CRM design}
\label{appendix:reproduction}
The reproduced outcome of the 7 scenarios presented in \citet{dehbi2021controlled} is in Table \ref{tbl:reproduce1} where the same skeleton of CRM is used.

\begin{table}[htbp]
\tiny
\begin{center}
\caption{Comparison of Reproduction of Backfill CRM }
\label{tbl:reproduce1}
\begin{tabular}{@{}ccccccccccccccc@{}}
\toprule
 & \multicolumn{7}{c}{Reproduced Backfill CRM}& \multicolumn{7}{c}{Backfill CRM} \\ \midrule
Dose level& 1  &2   & 3 & 4 &5  & 6 &7& 1  &2   & 3 & 4 &5  & 6 &7  \\  \midrule
\textbf{Scenario A}&   &   &  &  &  &  & &   &   &  &  &  &  &  \\
Efficacy probability& 0.05 & 0.15  &0.25  & 0.25 &0.25  &0.25  &0.25  &0.05 & 0.15  &0.25  & 0.25 &0.25  &0.25  &0.25  \\
Toxicity probability& 0.01& 0.04  &0.08  & 0.16 &0.25  &0.35  &0.46  & 0.01& 0.04  &0.08  & 0.16 &0.25  &0.35  &0.46 \\
\% of OBD selection& 0\%& 0.3\% &\textbf{29.2\%}  &47.2\%  & 17.4\% & 5.5\% &0.4\%& 0\%& 0\% &\textbf{29\%}  &46\%  & 16\% & 8\% &1\%  \\
\% of MTD selection&  0\%& 0\% & 2.0\%  & 26.9\%  & 52.0\% & 17.6\% &1.5\%    &   &  &  &  &  & &\\
Efficacy estimation&  0.15 & 0.17  &0.20  &0.23  &0.25  &0.28  & 0.29  &   &  &  &  &  & & \\ 
Number of patients enrolled & 8.9& 10.4 & \textbf{11.1} & 11.9 & 9.2 & 4.3 & 1.3& 9.1& 10.3 & \textbf{11.4} & 10.8 & 9.7 & 5.1 & 1.1\\
Number of patients backfill & 5.8& 7.3& 7.1 & 4.9 &1.7 &0.3 &0  &   &  &  &  &  & &\\
\textbf{Scenario B}&   &   &  &  &  &  &  &   &   &  &  &  &  & \\
Efficacy probability& 0.05 & 0.1  &0.15  & 0.15 &0.15  &0.15  &0.15 & 0.05 & 0.1  &0.15  & 0.15 &0.15  &0.15  &0.15  \\
Toxicity probability& 0.01& 0.04  &0.08  & 0.16 &0.25  &0.35  &0.46& 0.01& 0.04  &0.08  & 0.16 &0.25  &0.35  &0.46  \\
\% of OBD selection& 0\%& 0.4\% & \textbf{28.2\%}  & 49.2\%  & 17.0\% & 4.7\% &0.5\% & 0\%& 0\% & \textbf{32\%}  & 46\%  & 15\% & 6\% &1\%\\
\% of MTD selection&  0\%& 0\% & 1.8\%  & 26.2\%  & 52.9\% & 17.8\% &1.3\%    &   &  &  &  &  & &\\
Efficacy estimation&  0.10 & 0.11  &0.13  &0.14  &0.15  &0.17  & 0.18  &   &  &  &  &  & & \\ 
Number of patients enrolled & 9.5& 10.6 & \textbf{10.8} & 11.4 & 9.1 & 4.3 & 1.3& 10.3& 10.8 & \textbf{10.3} & 10.3 & 9.7 & 4.6 & 1.1\\
Number of patients backfill & 6.4& 7.5& 6.9 & 4.4 &1.6 &0.3 &0  &   &  &  &  &  & &\\
\textbf{Scenario C}&   &   &  &  &  &  &  &   &   &  &  &  &  & \\
Efficacy probability& 0.04 & 0.08  &0.12  & 0.16 &0.2  &0.24  &0.28  & 0.04 & 0.08  &0.12  & 0.16 &0.2  &0.24  &0.28 \\
Toxicity probability& 0& 0  &0.01  & 0.04 &0.08  &0.16  &0.25 & 0& 0  &0.01  & 0.04 &0.08  &0.16  &0.25  \\
\% of OBD selection& 0\%& 0\% & 0\%  & 25.1\%  & 23.0\% & 4.9\% & \textbf{47.0\%} & 0\%& 0\% &0\%  &30\%  & 10\% & 9\% &\textbf{51\%}  \\
\% of MTD selection&  0\%& 0\% & 0\%  & 0.1\%  & 0.7\% & 6.6\% &92.6\%   &   &  &  &  &  & & \\
Efficacy estimation&  0.10 & 0.12  &0.14  &0.16  &0.19  &0.21  & 0.24  &   &  &  &  &  & & \\ 
Number of patients enrolled & 8.3 & 8.0 & 8.1 & 7.7 & 7.4 & 6.8 & \textbf{10.8}& 9.7 & 9.1 & 8.0 & 7.4 & 6.3 & 5.7 & \textbf{10.9}\\
Number of patients backfill & 5.3& 5.0& 5.0 & 4.6 &4.1 &2.9 &0  &   &  &  &  &  & &\\
\textbf{Scenario D}&   &   &  &  &  &  & &   &   &  &  &  &  &  \\
Efficacy probability& 0.07 & 0.14  &0.21  & 0.28 &0.35  &0.42  &0.49& 0.07 & 0.14  &0.21  & 0.28 &0.35  &0.42  &0.49  \\
Toxicity probability& 0& 0  &0.01  & 0.04 &0.08  &0.16  &0.25& 0& 0  &0.01  & 0.04 &0.08  &0.16  &0.25  \\
\% of OBD selection& 0\%& 0\% &0.2\%  &12.2\%  & 19.7\% & 8.7\% & \textbf{59.2\%}& 0\%& 0\% &0\%  &14\%  & 13\% & 13\% &\textbf{60\%}   \\
\% of MTD selection&  0\%& 0\% & 0\%  & 0\%  & 0.8\% & 6.6\% &92.6\%    &   &  &  &  &  & &\\
Efficacy estimation&  0.15 & 0.18  &0.23  &0.28  &0.33  &0.39  & 0.44   &   &  &  &  &  & &\\
Number of patients enrolled & 8.0 & 8.0 & 7.9 & 7.8 & 7.6 & 6.9 & \textbf{10.8}& 8.6 & 8.6 & 8.0 & 7.4 & 6.8 & 6.3 & \textbf{10.8}\\
Number of patients backfill & 5.0& 5.0& 4.9 & 4.7 &4.3 &3.1 &0  &   &  &  &  &  & &\\
\textbf{Scenario E}&   &   &  &  &  &  &  &   &   &  &  &  &  & \\
Efficacy probability& 0.05 & 0.1  &0.15  & 0.2 &0.2  &0.2  &0.2 & 0.05 & 0.1  &0.15  & 0.2 &0.2  &0.2  &0.2 \\
Toxicity probability& 0.04& 0.08  &0.16  & 0.25 &0.35  &0.46  &0.56 & 0.04& 0.08  &0.16  & 0.25 &0.35  &0.46  &0.56  \\
\% of OBD selection& 0\%& 3.7\% &47.1\%  & \textbf{38.7\%}  & 9.9\% & 0.6\% &0\% & 0\%& 5\% &45\%  &\textbf{36\%}  & 11\% & 2\% &0\%  \\
\% of MTD selection&  0\%& 1.6\% & 26.1\%  & 49.0\%  & 21.3\% & 2.0\% &0\%   &   &  &  &  &  & & \\
Efficacy estimation&  0.10 & 0.12  &0.14  &0.17  &0.20  &0.22  & 0.25   &   &  &  &  &  & &\\
Number of patients enrolled & 12.0& 12.8 & 13.3 & \textbf{11.1} & 5.8 & 1.7 & 0.3 & 12.0 & 12.5 & 13.7 & \textbf{10.8} & 5.7 & 1.7 & 0.6\\
Number of patients backfill & 8.5& 8.4& 6.3 & 3.0 &0.7 &0.1 &0  &   &  &  &  &  & &\\
\textbf{Scenario F}&   &   &  &  &  &  &  &   &   &  &  &  &  & \\
Efficacy probability& 0.04 & 0.08  &0.12  & 0.16 &0.2  &0.24  &0.24& 0.04 & 0.08  &0.12  & 0.16 &0.2  &0.24  &0.24  \\
Toxicity probability& 0.04& 0.08  &0.16  & 0.25 &0.35  &0.46  &0.56 & 0.04& 0.08  &0.16  & 0.25 &0.35  &0.46  &0.56  \\
\% of OBD selection& 0\%& 3.8\% &48.2\%  &\textbf{37.7\%}  & 9.2\% & 1.1\% &0\% & 0\%& 6\% &45\%  &\textbf{34\%}  & 13\% & 3\% &0\%  \\
\% of MTD selection&  0\%& 1.7\% & 25.9\%  & 50.5\%  & 19.5\% & 2.3\% &0.1\%    &   &  &  &  &  & &\\
Efficacy estimation&  0.08 & 0.10  &0.12  &0.14  &0.17  &0.20  & 0.22  &   &  &  &  &  & & \\
Number of patients enrolled & 11.9& 12.7 & 13.4 & \textbf{11.0} & 6.0 & 1.7 & 0.3& 12.0& 12.5 & 13.1 & \textbf{10.8} & 6.3 & 1.7 & 0.6\\
Number of patients backfill & 8.4& 8.3& 6.3 & 3.0 &0.8 &0.1 &0  &   &  &  &  &  & &\\
\bottomrule
\end{tabular}
\end{center}
\end{table}

\clearpage

\section{Comparison with the mTPI-2 design given $p_T=0.25$  }
\label{appendix:com_mTPI2}
\begin{figure}[!htbp]
    \centering
    \includegraphics[width=1.0\textwidth]{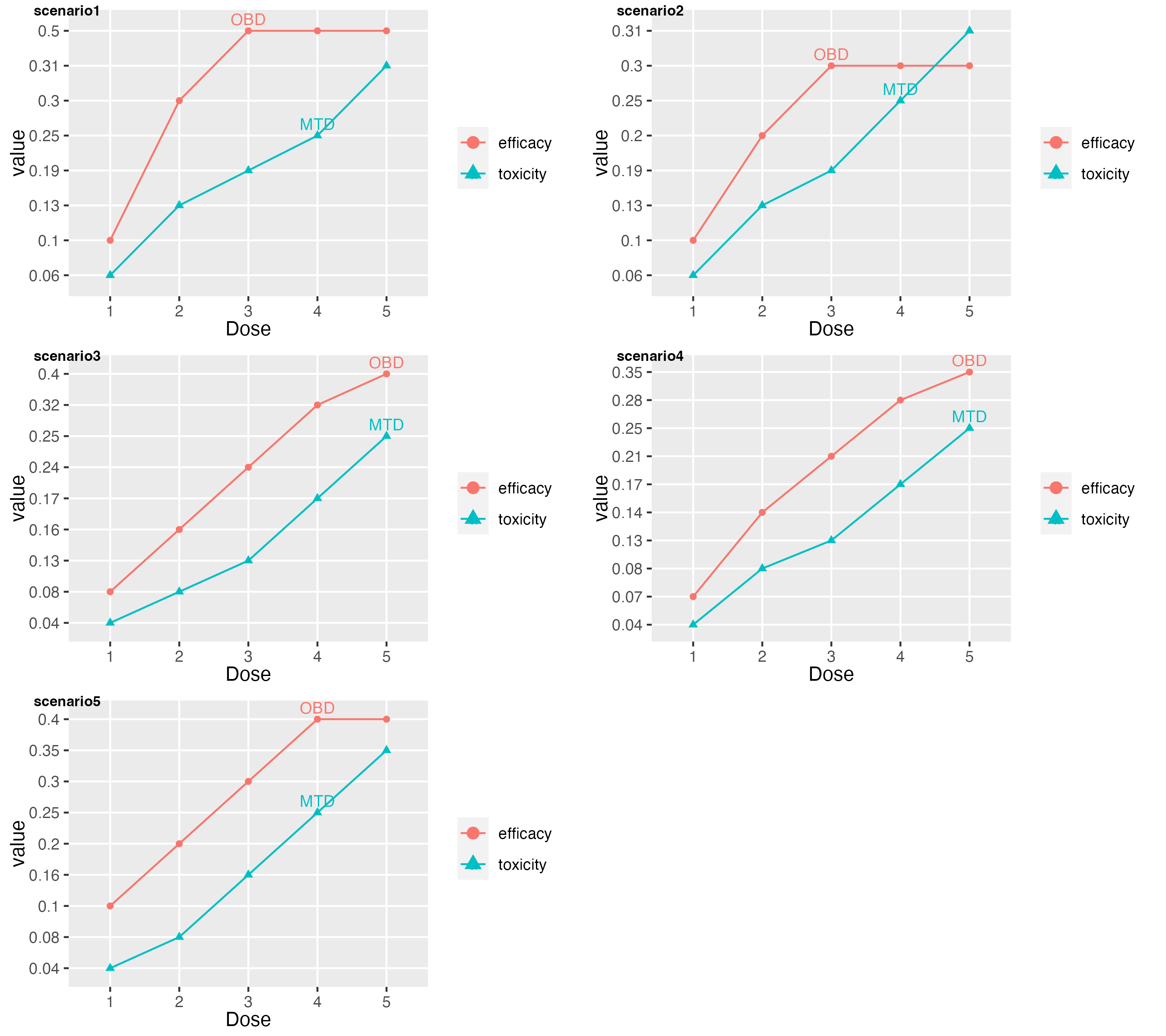}
    \caption{  The five simulation scenarios with target toxicity probability of 0.25. The (MTD, OBD) are (4,3), (4,3), (5,5), (5,5), and (4,4) for scenarios 1-5, respectively.   }
    \label{fig:5dose-1}
\end{figure}

\begin{table}[!htbp]
\tiny
\begin{center}
\caption{Comparison of Bi3+3 and mTPI-2 in five scenarios. The target toxicity probability is 0.25, EI =[0.2,0.3], sample size is 42.4 for Bi3+3 and 43.0 for mTPI-2.}
\label{tbl:oc_vs_mtpi2_pt0.25}
\begin{tabular}{@{}ccccccccccc@{}}
\toprule
 & \multicolumn{5}{c}{Bi3+3} & \multicolumn{5}{c}{mTPI-2} \\ \midrule
Dose level&  1&2  &3  &4  &5 &  1&2  &3  &4  &5 \\ \midrule
\textbf{Scenario 1}&  &  &  &  & &  &  &  &  &  \\
Efficacy probability& 0.1 & 0.3 &0.5  &0.5  &0.5 &  &  &  &  & \\
Toxicity probability& 0.06 &0.13  & 0.19  &0.25  & 0.31& 0.06 &0.13  & 0.19  &0.25  & 0.31 \\
\% of OBD selection& 3.1\%  & 34.4\% & \textbf{34.2\%}  &19.0\%  &9.3\% &   &  &   &  &  \\
\% of MTD selection& 3.1\%  & 20.2\% & 32.4\%  &\textbf{28.3\%}  &16.0\%& 4.6\% & 26.6\% &  37.2\% & \textbf{23.3\%} & 8.3\% \\
Efficacy estimation& 0.20 &0.31  &0.43  &0.49  &0.53 &  &  &  &  &  \\ 
Number of patients enrolled & 9.1 &10.3 & \textbf{10.3} & \textbf{6.6} &4.2& 6.4& 12.3 & 12.6& \textbf{7.8} & 3.8\\
Number of patients backfill &4.4 &2.9 &2.2 &0.9 &0& & & & &\\
Trial Duration& \multicolumn{5}{c}{491}  & \multicolumn{5}{c}{600}\\
\textbf{Scenario 2}&  &  &  &  & &  &  &  &  &  \\
Efficacy probability& 0.1 &0.2  & 0.3 &0.3  &0.3 &  &  &  &  &  \\
Toxicity probability&0.06 &0.13  & 0.19  &0.25  & 0.31& &  &   &  &  \\
\% of OBD selection& 3.1\%  & 28.4\% &\textbf{32.3\%}  & 23.3\% &12.9\% &   &  & &  & \\
\% of MTD selection& 3.1\% & 18.9\% & 33.6\% & \textbf{28.8\%} & 15.6\% &  &  &  &  &  \\
Efficacy estimation& 0.16 &0.20  &0.26  &0.31  &0.34 &  &  &  &  &  \\ 
Number of patients enrolled & 10.0 & 10.9 & \textbf{10.2} & \textbf{6.6} & 4.2&  &  &  &  & \\
Number of patients backfill & 5.3&3.5 &2.1 &0.9 &0& & & & &\\
\textbf{Scenario 3}&  & &  &  &  &  & &  &  &\\
Efficacy probability& 0.08 &0.16  &0.24  &0.32  &0.4&  &  &  &  &   \\
Toxicity probability& 0.04 &0.08  &0.13  &0.17  &0.25& 0.04 &0.08  &0.13  &0.17  &0.25  \\
\% of OBD selection& 0.3\% & 10.3\% & 20.5\% & 35.0\% & \textbf{33.9\%} &  &  &  &  & \\
\% of MTD selection& 0.3\% & 5.6\% & 17.6\% & 37.9\% & \textbf{38.6\%} & 0.6\% & 9.7\% & 25.4\% & 37.2\% & \textbf{27.1\%}\\
Efficacy estimation& 0.14 & 0.18 &0.24  & 0.30 &0.35&  &  &  &  &  \\ 
Number of patients enrolled& 8.6 & 8.9 & 9.4 & 8.3 & \textbf{8.3}& 4.5 & 7.7 & 10.6 & 11.0 & \textbf{9.3}\\
Number of patients backfill &5.0 &3.9 &2.9 &1.7 &0 & & & & & \\
Trial Duration& \multicolumn{5}{c}{495}  & \multicolumn{5}{c}{608}\\
\textbf{Scenario 4}&  &  &  &  & &  &  &  &  &  \\
Efficacy probability& 0.07 &0.14  &0.21  &0.28  &0.35 &  &  &  &  &  \\
Toxicity probability& 0.04 &0.08  &0.13  &0.17  &0.25 &  & &  &  &  \\
\% of OBD selection& 0.3\% & 9.7\% &20.4\%  & 34.7\% & \textbf{34.9\%} &  &  &  &  & \\
\% of MTD selection& 0.3\% & 5.6\% & 16.4\% & 37.6\%  & 40.1\% &  &  &  &  & \\
Efficacy estimation&  0.13& 0.16 &0.21  &0.26  &0.31 &  &  &  &  &   \\ 
Number of patients enrolled& 8.8 & 8.9 & 9.5 & 8.3 & \textbf{8.4}&  &  &  &  & \\
Number of patients backfill & 5.1&3.9 &3.1 &1.7 &0& & & & & \\
\textbf{Scenario 5}&  &  &  &  & &  &  &  &  & \\
Efficacy probability& 0.1& 0.2  &0.3  &0.4  &0.4 &  &  &  &  &  \\
Toxicity probability& 0.04 &0.08  &0.16  &0.25  &0.35 & 0.04 &0.08  &0.16  &0.25  &0.35 \\
\% of OBD selection& 0.3\% & 19.5\% & 38.3\% & \textbf{30.6\%} & 11.3\%&  &  &  &  &  \\
\% of MTD selection& 0.3\% & 10.8\% & 39.7\% & \textbf{35.7\%} & 13.5\% & 0.7\% & 15.3\% & 44.6\% & \textbf{33.2\%} & 6.2\% \\
Efficacy estimation& 0.16 & 0.22 &0.29  &0.36  &0.41 &  &  &  &  &  \\
Number of patients enrolled & 8.6 & 9.4 & 11.2 & \textbf{8.2} & 4.9& 4.5 & 9.0 & 14.7 & \textbf{10.6} & 4.2\\
Number of patients backfill &4.9 &3.7 &2.4 &1.0 & & & & & &\\
Trial Duration& \multicolumn{5}{c}{491}  & \multicolumn{5}{c}{600}\\\bottomrule
\end{tabular}
\end{center}
\end{table}

\end{appendices}

\end{document}